\documentclass{ws-procs975x65}

\begin{document}

\title{TRANSITS OF VENUS AND THE ASTRONOMICAL UNIT: FOUR CENTURIES OF INCREASING PRECISION}

\author{C. SIGISMONDI,$^1,^2$}

\address{$^1$ICRANET,Sapienza University and 
Pontifical Atheneum Regina Apostolorum, Rome, Italy sigismondi@icra.it 
$^2$Observatorio Nacional, Rio de Janeiro, Brazil}

\begin{abstract}
Only seven transits of Venus have been observed and studied up to now since the first astronomical observations with the telescope: 1639, 1761-69, 1874-82 and 2004-12. The measurement of the Astronomical Unit has been one of the main goal of  the study of this rare phenomenon, as well as the identification of the planetary atmosphere and its properties, the understanding of the black drop phenomenon and the measurement of the solar diameter. The 1639 transit observed by J. Horrocks is presented in more detail.
\end{abstract}


\bodymatter

\section{Introduction}

The transits of Venus have been observed and studied after Johannes Kepler's predictions published in the Tabulae Rudolphinae (1627). On November 7, 1631 Pierre Gassendi observed firstly a planet transiting over the Sun: Mercury. The following transit of Venus was not seen. In 1639 Jeremiah Horrocks observed the first stages of the transit of Venus on December 4 (Gregorian Calendar), and used this observation to determine the distance Sun-Earth. 
Later in 1716 Edmond Halley proposed the method to exploit the transit of Venus to measure the distance Sun-Earth $D\odot\oplus$, and for the 1761-1769 both France and United Kingdom organized worldwide expedition (Jean-Baptiste Chappe d'Auteroche, Alexandre Guy Pingr\'e, James Cook and Charles Green) to observe the transits of Venus and to measure $D\odot\oplus$.
This parameter was also the goal of the next pair of transits in 1874 and 1882, with the help of photography and spectroscopy. 
In 2004 there was a great didactic interest,\cite{zero} being the transit entirely visible from all Europe.
Moreover the transits of 2004-12 have been studied in view of exoplanets with atmosphere in transits over their stars, 
as in the Venus Twilight Experiment.\cite{tanga} 
Venus and Mercury transits are also crucial to know the past history of the solar diameter. Through the W parameter, $W=dln R / dln L$ the logarithmic derivative of the radius with respect to the luminosity, the past values of the solar luminosity can be recovered. The black drop phenomenon affects the evaluation of the instants of internal and external contacts between the planetary disk and the solar limb.\cite{cinque} With these observed instants compared with the ephemerides the value of the solar diameter is recovered. 
The expected accuracy on the diameter of the Sun measured from ground with this method cleaned from black drop\cite{sigi0} is expected to be better than 0.01 arcsec,\cite{sigi} with good images of the ingress and of the egress taken each second, as the observations made in Chinese solar observatories of Huairou and Xi Chong during the observational champaign of 2012.

\section{The Astronomical Unit during the last four centuries}

The measurement of $D\odot\oplus$ is crucial for understanding the dimensions of our Universe, being the first step of the cosmological distance ladder. Stellar parallaxes, firstly measured in 1838 by Wilhelm Bessel, have the basis in the Astronomical Unit, the semiaxis major of the Earth's orbit.
The form of the Earth's orbit was precisely known after Kepler's Astronomia Nova (1609), 
and the value of the eccentricity (the double) was already known by Ptolemy (150 AD) after the data on the durations of the seasons.
Giovanni Domenico Cassini confirmed the Keplerian theory in 1655 with its "heliometer", a pinhole\footnote{It is interesting considering that these measurements were done with a method exploited to assess the variation of the obliquity of the Earth's axis already before the invention of the telescope. Paolo Toscanelli in 1475 built a similar instrument in Santa Maria del Fiore in Florence, and pope Clement XI ordered Francesco Bianchini to build one in Santa Maria degli Angeli (Rome) in 1702.} with meridian line located in the Basilica of San Petronio in Bologna. The angular variations of the solar diameter along the year confirmed the hypothesis of the bisection of the eccentricity.\cite{heilbron}
Jeremiah Horrocks (1618-1641), a genial student of Cambridge, recalculated himself the ephemerides of Venus being in disagreement with the ones of Philip Lansberge (1561-1632), and prepared accurately the observation of the transit on the Sunday 24 November 1639 (December 4 of Gregorian Calendar). He prudently begun the observations to the day before, in order to take into account for errors, and he was enough lucky to observe the beginning of the transit occurring one hour before the local sunset of the predicted day. His observation was published 21 years after his death by Johannes Hevelius in 1662. Horrocks started to study the ephemerides when he was 14-15 years old, observed the transit at 21, and died in his 22nd year.\footnote{"Jeremiah Horrox possessed one of the most original minds of the seventeenth century. A follower of Tycho, Horrox combined a gift for instrumentation with a theoretical genius that later won the acclaim of Hevelius and Flamsteed. His contributions to the lunar theory were the first significant advances on the subject since antiquity, and earned the praise of Newton himself in the pages of Principia."\cite{uno}}
Horrocks wrote to his friend William Crabtree (1610-1644) on 5th November 1639 his expectations on the transit.
\footnote{"The reason why I am writing you now is to inform you of the extraordinary conjunction of the Sun and Venus which will occur on November 24. At which time Venus will pass across the Sun. Which, indeed, has never happened for many years in the past nor will happen again in this century. I beseech you, therefore, with all my strength, to attend to it diligently with a telescope and to make whatever observation you can, especially about the diameter of Venus, which, indeed, is 7' according to Kepler, 11' according to Lansberge, and scarcely more than 1' according to my proportion."\cite{due}}
According to Gassendi the angular diameter of Mercury was 20 arcsec, much smaller than Kepler beliefs. Horrocks could state that Venus angular diameter was 76 arcsec, again much smaller than all previous hypotheses.

The consideration that the angular diameter of the Earth and Venus would have been of the same order of magnitude as seen from the Sun led Horrocks to formulate the hypothesis that the distance Earth-Sun was 95.7 millions Km, the largest ever thought up to his times. 
Both Horrocks and Gassendi made two measurements in excess with respect to the real values, because of diffractive and chromatic effects of the point-spread function of their small refracting telescopes.
Halley proposed to use the transits of Venus of 1761-69 to measure accurately the Astronomical Unit with the method of parallaxes which nowadays is classic. And the values of these measurements started to converge to the actual value of 149.6 millions Km. \cite{quattro}
The Astronomical Unit was perfectly known already in 1874 and 1882 transits 
as reading the reports of S. Newcomb:\cite{tre} the solar parallax was measured from the contact timings of the transits of Venus as $8.794\pm0.018$" 
being $8.794148\pm0.000007$" the current value adopted in the astronomical constants. The Astronomical Unit is the fundamental distance for all astronomy and astrophysics 
and for gravitational physics to determine the product $G \times M_\odot$ using the Newton and Kepler's laws.

\begin{figure}[t]
\psfig{file=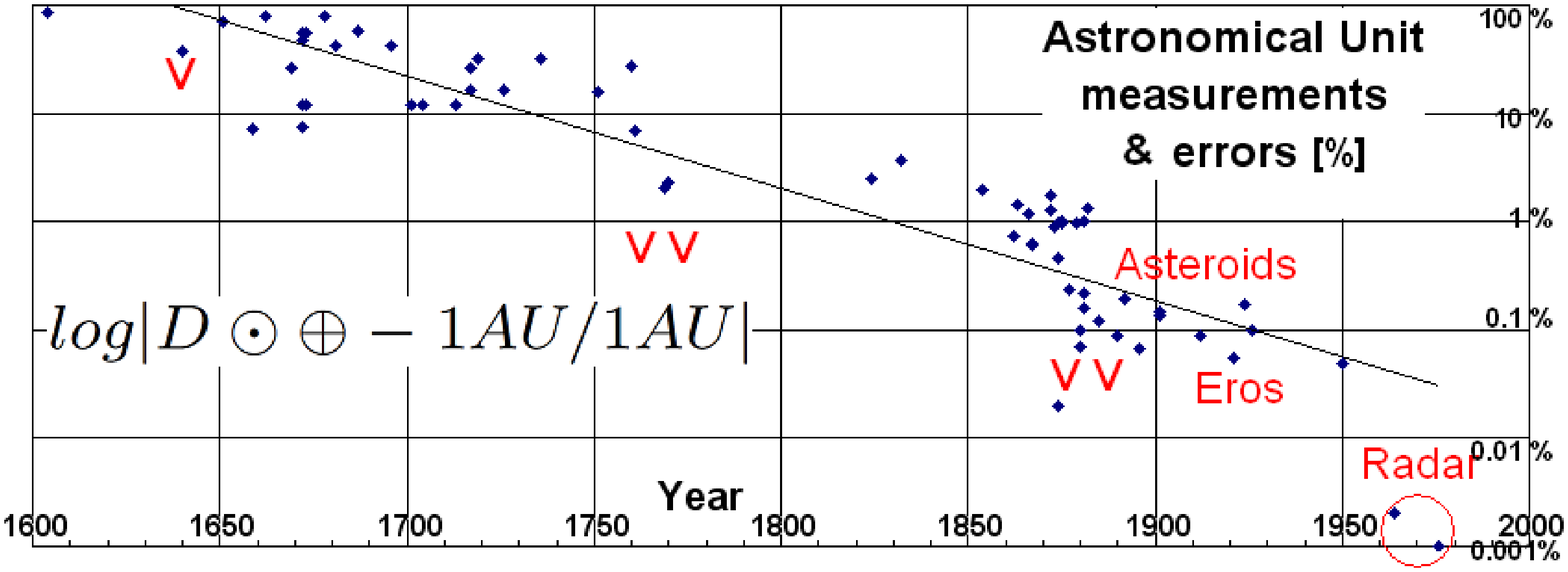,width=5in}
\caption{The improvement on the measurements of the distance Sun-Earth $D\odot\oplus$ from 1600 to present adapted from table 2 of D. Hughes.\cite{quattro} Venus transits are marked with V. The percentual errors obtained with $log|D\odot\oplus-1AU/1AU|$ are indicated on the right side. The linear fit is in agreememnt with the trend of accuracy in astronomical angular measurements studied by Pledge (1939:291),\cite{Pledge} and with the one of time accuracy published by Howse.\cite{Howse}}
\label{aba:fig1}
\end{figure}




\end{document}